\documentclass[twocolumn]{aastex62}
\usepackage{amssymb}
\usepackage{amsmath,color}
\usepackage{epsfig}
\usepackage{graphicx}
\usepackage{bbm}
\usepackage{subfigure}
\usepackage{multirow}
\usepackage{verbatim}
\usepackage[T1]{fontenc}
\usepackage[utf8]{inputenc}
\usepackage[normalem]{ulem}
\usepackage{longtable}

\usepackage{url}
\usepackage{indentfirst}

\usepackage{color}
\usepackage{ulem}

\begin{document}

\title{New probe of gravity: strongly lensed gravitational wave multimessenger approach}

\correspondingauthor{Bin Hu}
\email{bhu@bnu.edu.cn}

\author{Tao Yang}
\author{Bin Hu}
\affiliation{Department of Astronomy, Beijing Normal University, Beijing, 100875, China}

\author{Rong-Gen Cai}
\affiliation{CAS Key Laboratory of Theoretical Physics, Institute of Theoretical Physics, Chinese Academy of Sciences, P.O. Box 2735, Beijing 100190, China}
\affiliation{School of Physical Sciences, University of Chinese Academy of Sciences, No. 19A Yuquan Road, Beijing 100049, China}

\author{Bin Wang}
\affiliation{Center for Gravitation and Cosmology, Yangzhou University, Yangzhou 225009, China}

\begin{abstract}
Strong gravitational lensing by galaxies provides us with a unique opportunity to understand the nature of gravity on galactic and extra-galactic scales.
In this paper, we propose a new multimessenger approach using data from both gravitational wave (GW) and the corresponding electromagnetic (EM) counterpart to infer the constraint of the modified gravity (MG) theory denoted by the scale dependent phenomenological parameter.
To demonstrate the robustness of this approach, we calculate the time-delay predictions by choosing various values of the phenomenological parameters and then compare them with that from the general relativity (GR).
For the third generation ground-based GW observatory, with one typical strongly lensed GW+EM event, and assuming that the dominated error from the stellar velocity dispersions is 5\%, the GW time-delay data can distinguish an 18\% MG effect on a scale of  tens of kiloparsecs with a $68\%$ confidence level. Assuming GR and a Singular Isothermal Sphere mass model, there exists a simplified consistency relationship between time-delay and imaging data. This relationship does not require for the velocity dispersion measurement, and hence can avoid major uncertainties. By using this relationship, the multimessenger approach is able to distinguish an $8\%$ MG effect. Our results show that the GW multimessenger approach can play an important role in revealing the nature of gravity on the galactic and extra-galactic scales.
\end{abstract}




\section{Introduction} \label{sec:intro}
Einstein's general relativity (GR) has been precisely tested on
solar system scales~\cite{Bertotti:2003rm,Shapiro:2004zz}, such as the Eddington's measurement of light deflection during the solar
eclipse of 1919~\cite{Dyson:1920cwa}; the observation of the gravitational
redshift~\cite{Pound:1960zz}; the successful operation of the Global Positioning Satellites~\cite{Ashby:2003vja}; the measurements of the Shapiro time-delay~\cite{Shapiro:1964uw}; and the verification of energy loss via
gravitational waves (GWs) in the Hulse--Taylor pulsar~\cite{Taylor:1979zz}. 
We refer to the living review \cite{Will:2014kxa} for the updated theoretical and observational progress in testing gravity.
However, the long-range nature of gravity on the extra-galactic scale is still poorly
understood. Testing gravity with higher accuracy has been continuously pursued over the past decades. The purpose of these activities is not only  to examine
a specific model, but also to reveal the nature of gravitational phenomena, such as dark matter and dark energy, on the cosmological scales.
The parameterized post-Newtonian (PPN) framework~\cite{Thorne:1970wv} provides us with a systematic way to quantify the deviation from GR.
The scale independent post-Newtonian parameter $\gamma_{\texttt{PPN}}$ represents the ratio between dynamical mass and lensing mass.
The former is the mass determining the motion of nonrelativistic objects, such as the stellar velocity dispersion, while the latter is the mass determining the path deflection of relativistic particles, such as the Einstein radius of strong lensing image and Shapiro time-delay.
GR predicts the equivalence of these two masses ($\gamma_{\texttt{PPN}}=1$), while the alternative models break it due to the extra scalar degree of freedom, which mediates the fifth force. Here we note that we do not consider the massive graviton or the anomaly of the speed of GW, which is tightly constrained by the recent GW detection of~\cite{Monitor:2017mdv}.

Strong gravitational lensings around galaxies provide us with a unique opportunity to probe modifications to GR over a range of redshifts and on/above kiloparsec scales~\cite{Bolton:2006yz,Smith:2009fn,Schwab:2009nz,Cao:2017nnq}.
Recently, \cite{Collett:2018gpf} estimated $\gamma_{\texttt{PPN}}$ to be $0.97\pm0.09$ at a $68\%$ confidence level by using
a nearby lens, ESO 325-G004. In these analysis, the dynamical mass is estimated by the spectroscopic measurement of the stellar velocity dispersion of the lens galaxy, and the lensing mass is reconstructed by the measurement of the Einstein radius. All of these data are obtained by using the electromagnetic signals.

The first GW event GW150914 from
the merger of the binary black hole, detected by the LIGO Scientific Collaboration~\footnote{\url{https://ligo.org}} and the Virgo Collaboration~\footnote{\url{http://www.virgo-gw.eu}}, opened a new observational window to explore our universe~\cite{Abbott:2016blz}.
Moreover, the following event, GW170817 from a binary neutron star combined with the electromagnetic (EM) counterpart, announced the beginning of the golden era of multimessenger astronomy~\cite{TheLIGOScientific:2017qsa}.
With the development of multimessenger observations, searching the strongly lensed GW event associated with the EM counterpart should be
attractive. Recently, several researches have investigated this issue for the LIGO events~\cite{Smith:2017mqu,Hannuksela:2019kle,Broadhurst:2019ijv}.
In the context of the Laser Interferometer Space Antenna (LISA) interferometric detector in space,~\footnote{\url{https://www.lisamission.org}}, strong lensing of LISA target sources (supermassive BHs) and its application on cosmology have also been discussed~\cite{Sereno:2010dr,Sereno:2011ty}.
The third generation of ground-based GW observatories, such as the Einstein Telescope~\cite{Punturo:2010zz}~\footnote{\url{http://www.et-gw.eu}}, is expected to detect $10^4-10^5$ events per year and $50-100$ among them  are  strongly lensed~\cite{Piorkowska:2013eww,Biesiada:2014kwa,Ding:2015uha}.
The redshift of the unlensed event is expected to reach around $z\sim 2$. While, due to the magnification by intervening galaxies, the redshift range would extended to $z\sim 8$~\cite{Biesiada:2014kwa}. The observation of a strongly lensed GW with an EM counterpart should be very promising.
The time delays of GW from different paths, are expected to be a few months; while the duration of each signal is less than $0.1$ s. 
The transient nature of this double compact object merger event would lead to a very accurate measurement of the time-delay compared to the traditional light-curve measurement in the optical domain ($\sim 3\%$); hence we can ignore this uncertainty of time-delay measurement from GW.
Furthermore, the lensing mass reconstruction from the EM counterpart images has a systematic uncertainty of around the $0.6\%$ level~\cite{Liao:2017ioi}.  While, for the traditional lensed quasar system, the uncertainties of the time-delay and lensing mass reconstruction are both of order $\mathcal{O}(3\%)$~\cite{Liao:2014cka,Suyu:2016qxx}. 
This improvement could be illustrated as follows.  For the optical quasar system, the Fermat potentials are recovered from the lensed host galaxy image. However, in the central part of the active galactic nucleus (AGN), due to the overly saturated exposure, even a tiny mismatch, when extracting the AGN images as the scaled point spread functions (PSFs), would lead to a nonnegligible discrepancy. 
\cite{2017MNRAS.465.4634D} carried out a data challenge and  found that even if the perfect PSF is given, a significant residual in the central AGN area is still inevitable. Fortunately, one does not encounter these difficulties while studying the lensed GW+EM events since these systems do not possess the bright point images.  ~\cite{Liao:2017ioi} performed the simulations and used state-of-the-art lens modelling techniques; they found that the relative uncertainty of the Fermat potential reconstruction could reach the $0.6\%$ level. This improves the traditional optical $3\%$ level by a factor of 5.
Recently, ~\cite{Liao:2017ioi} forecasted the robustness of Hubble parameter estimation by using the strongly lensed GW+EM signals. They found that $10$ such systems are able to provide a Hubble constant uncertainty of $0.68\%$ for a flat $\Lambda$ cold dark matter ($\Lambda$CDM) universe by using the Einstein Telescope. 
The possibility of  to exploring the GW speed anomaly was investigated by \cite{Fan:2016swi}.

Using the GW+EM multimessenger approach to test GR has been investigated by many studies~\cite{Baker:2017hug,Abbott:2018lct,Nishizawa:2019rra}. The strongly lensed of GW+EM should also provide us with a very promising tool to test GR. The optical strong lensing test of GR under the PPN framework has been proposed in~\cite{Bolton:2006yz,Smith:2009fn,Schwab:2009nz,Cao:2017nnq}. 
All of these studies are limited only in the EM domain, which is out of date
in the richness of GW+EM multimessenger observation today. Having the significant benefits from strongly lensed GW addressed above,
it inspired us to consider the strongly lensed GW+EM system as a new probe to the nature of gravity on kiloparsec scales.

\section{Model Setup} \label{sec:model}

In the limit of a weak gravitational field, consider the perturbed
Friedmann-Lema\^ itre-Robertson-Walker (FLRW) metric in the conformal Newtonian gauge
\begin{equation}
ds^2=-\left(1+\frac{2\Psi}{c^2}\right)c^2 dt^2+a^2\left(1-\frac{2\Phi}{c^2}\right)d\vec{x}^2\,,
\label{eq:ds}
\end{equation}
where $a$ is the background scale factor, $\Psi$ and $\Phi$ are the Newtonian potential
and spatial curvature perturbation. Non-relativistic particle motion only responds to a spatial gradient of
the Newtonian potential $\Psi$, due to the fact that its velocity is much less than the  speed  of light $c$.
While a relativistic particle is affected by the gradient of the Weyl potential, $\Phi_+=(\Phi+\Psi)/2$.
GR predicts the equivalence of Newtonian and Weyl potentials, or the equivalence of Newtonian potential $\Psi$ and the spatial curvature perturbation $\Phi$, namely $\gamma_{\texttt{PPN}}\equiv\Phi/\Psi=1$.
In contrast, the alternative models of gravity typically contain an additional scalar
degree of freedom that mediates the fifth force, and hence breaks this degeneracy.

On kiloparsec scales, strong gravitational lensing, combined with stellar
kinematics of the lens, allows a test of the weak field metric of gravity.
Measurements of the stellar velocity dispersion determine
the dynamical mass, whereas measurements of the image positions of the lens and time-delay between the multiple images determine the lensing mass.
They reflect the nature of Newtonian and Weyl potentials of the underlining gravity, respectively. In GR, the two potentials are the same, but many alternative
 gravity models predict the ratio of the two potentials to be scale dependent. For example, for the generic dark energy/modified gravity (MG) models, the unified parametrization of linear perturbations under the quasi-static approximation is given by~\cite{Pogosian:2016pwr}. On the nonlinear scale, a parametrization of MG is given by~\cite{Lombriser:2016zfz}, which unifies different screening mechanisms. All of these indicate that MG effect is a scale dependent phenomenon. There exist a plenty of parameter spaces to allow a spatial scale dependent modifications to standard gravity.
However, the analysis made in references~\cite{Bolton:2006yz,Smith:2009fn,Schwab:2009nz,Cao:2017nnq,Collett:2018gpf} assumed a constant PPN parameter over the length scales relevant for their studies. In this paper, we directly parameterize the Weyl potential, which is the most relevant parameter to the lensing.  As we know, the viable models of MG have to shield the fifth force under a certain scale, namely the screening scale~\cite{Joyce:2014kja}. The MG effect only arises above these scales.

In this paper, we introduce a scale dependent phenomenological parameter to denote the deviation of the Weyl potential from the Newtonian one, that is, $\Sigma(r)=\Phi_+/\frac{-GM}{r}$ in real space. 
Generally speaking, the MG effect shall also depend on the morphology of environment of the galaxies. However, most of the studies of screening mechanism in the current literature (see~\cite{Koyama:2015vza} for a review) are assuming spherical symmetry. This is due to the fact that,  on the one hand, spherical symmetry is computationally feasible; and, on the other hand, the nonsphericity, such as ellipticity, is not the major issue in the MG modelling for now. Our current galaxy survey data quality cannot tell us more than the spherical model in the gravity test. As for our studies, we explicitly assumed that the size of the screening scale of the MG effect is larger than the disk size of the lens galaxy. Hence, the symmetry of the lens is irrelevant to its environment (screening scale).

The reason we want to test the scale dependence of the $\Sigma$ function is the following. As shown in Fig. 1 of~\cite{Pogosian:2016pwr}, the scale dependence of the $\Sigma$ function is crucial for the gravity test. 
Take the generalized Brans--Dicke model as an example, which includes $f(R)$ gravity~\cite{Capozziello:2003tk,Carroll:2003wy,Appleby:2007vb,Hu:2007nk,Starobinsky:2007hu}, chameleon~\cite{Khoury:2003aq}, symmetron~\cite{Hinterbichler:2010es} and dilaton~\cite{Damour:1994zq,Brax:2011ja} models. The extra scalar field could give rise to a sizeable modifications to the Newtonian potential ($\mu$) below its Compton wavelength. But the modifications to Weyl potential ($\Sigma$) are usually tiny. As shown in Eq. (21) of~\cite{Pogosian:2016pwr}, in some subset of~\cite{Horndeski:1974wa} models, Weyl potential can significantly deviate from unity on the cosmological scale. Moreover, ~\cite{Lombriser:2016zfz} proposed a unified parameterization of the screening mechanism. Combining them, we can model the scale dependent MG dynamics from the linear down to the nonlinear regimes. Hence, if a sizeable deviation of $\Sigma$ from unity was detected, we can not only rule out GR, but also all kinds of generalized Brans--Dicke models. 
In addition, a scale dependent MG effect would be very helpful to break the degeneracy between the dynamical mass and the MG effect especially when we have information of both the time-delay and image positions of many lensed events in the future. 
Thus the motivation of our paper should be stated clearly here. We want to provide a new approach with the the exciting and fashionable strongly lensed multimessenger to test the GR on kiloparsec (galaxy) scales.

The photons and gravitons emitted from the source galaxies cumulate the MG effect along the line of sight when they approach to us.
Analogous to the parametric forms of $\mu$ and $\gamma$~\cite{Bertschinger:2008zb,Hojjati:2012rf,Pogosian:2016pwr}, which denote for the modifications to $\Psi$ and $\Phi$, we set the
parametric form of $\Sigma(r)$ as
\begin{equation}
\Sigma(r)=
\begin{cases}
1\,, & 0<r<r_{01} \\
\frac{1+\alpha_1\left(\frac{r-r_{01}}{w_{01}}\right)^2}{1+\alpha_2\left(\frac{r-r_{01}}{w_{01}}\right)^2}\,, & r\geq r_{01}\,,
\end{cases}
\label{eq:phi}
\end{equation}
where $\alpha_1$/$\alpha_2$ describes the magnitude of the modification of gravitational coupling, $r_{01}$
denotes the transition scale, and
$w_{01}$ represents the width of the transition step. 
We should note that the symbol $\Sigma$ here should not be confused with the surface density of the lens galaxy, which we usually adopted. In the following, we will use another symbol to denote the surface density parameter.
The typical Einstein radius of the galaxy lens is around $10$ kpc. We can imagine that the galaxy lensing data shall be sensitive to the MG effect which arises on the similar scales.
In this paper, we choose two specific transition scales, $10$ and $20$ kpc for comparison.
This means that the MG effect vanishes below the galactic scale.
A bonus is that we can adopt the standard lens modelling, such as SIS, without modifying the complicated galaxy dynamics.
More specifically, as we know, the scale above kiloparsecs to Megaparsecs is a blank window for gravity tests (precise tests)~\cite{Koyama:2015vza}. In this paper, we are not aiming to test a specific model, but using a phenomenological parameterization, which may cover a rather large model space. The two transition scales we give here are for comparison to show the robustness of this GW+EM time delay approach for the gravity test.
Hence, the MG effect comes into this system via a very clean way, namely the line-of-sight integral.
Take the deflection angle as an example. For a point mass lens, the MG effect 
contributes to two aspects. The first term of Eq. (\ref{eq:alpha}), which is similar to $\gamma_{\texttt{PPN}}$, corresponds to the enhancement/suppression of gravitational coupling of the relativistic species.  
The second term is due to the spatial variation of the gravitational coupling, which is absent in the constant $\gamma_{\texttt{PPN}}$ case.
Notice that differential measurement of $dz$ here is not redshift, but the Cartesian coordinate infinitesimal increment along the ``$z$'' direction (line of sight)
\begin{equation}
|\hat{\vec{\alpha}}| = \frac{4GM}{c^2}b\int^{\infty}_{0}\left(\frac{\Sigma}{r^3}-\frac{\Sigma_{,r}}{r^2}\right)dz\,.
\label{eq:alpha}
\end{equation}
Set $\mathcal{I}(b)=\int^{\infty}_{0}\left(\frac{\Sigma}{r^3}-\frac{\Sigma_{,r}}{r^2}\right)dz$, the defection angle can be written as
\begin{equation}
|\hat{\vec{\alpha}}| = \frac{4GM}{c^2}b \mathcal{I}(b)\,.
\label{eq:alphab}
\end{equation}
In GR, $\mathcal{I}(b)=1/b^2$, thus the deflection angle should be recovered to $\frac{4GM}{c^2b}$.

Given the lensing model and gravity theory, we can derive the multiple image positions and the time delay. The modified gravity theory with $\Sigma$ deviating from unity shall lead to a modification of the deflection angle.
For a general lens model, the deflection angle is, basically, the integration of the point mass over the corresponding lensing mass distribution. Hence, for any specific lens model with a given mass profile, we can calculate the deflection angle for a given impact parameter $b$.
Unlike the GR case, the time-delay in the MG case does not only depend on the geometric term, but also the effective lensing potential.  
The deviation of $\Sigma$ from unity represents the size of the MG effect. It produces both modifications to deflection angle and effective lensing potential, hence the final anomaly of time delay. After propagating all of the systematic uncertainties to the errors of time delay, we can compare the anomaly of time delay caused by MG to the systematic uncertainties to see how precisely we can distinguish MG theory from GR.  

The time delays from GW are the ``multiple sounds'' and the multiple images from EM counterpart are the ``multiple images''.
For a given system, the ``multiple sounds'' and ``multiple images'' can be derived simultaneously and respond to the same lensing mass.
Thus, if we could measure the dynamical mass (via velocity dispersions) accurately, the constraint on MG is straightforward.
Both of the time delay and image positions can be used, respectively, to estimate the difference between the lensing and dynamical mass.
In our paper, we choose the time delay from GW as the first part of our work in Sec.~\ref{sec:timedelay} to infer the constraints of the MG parameters. However, the image information from the EM counterpart should also be used as a supplement to improve the constraints, especially for a scale dependent MG effect. This is the second part of our work in Sec.~\ref{sec:test}.

\section{Time delay from GW} \label{sec:timedelay}
For a given lensed GW+EM event by a galaxy with a specific mass profile, the time delay between two images is given
by
\begin{equation}
\Delta t_{i,j}=\frac{1+z_l}{c}\frac{D_l D_s}{D_{ls}}\Delta \phi_{i,j}\,,
\label{eq:dt}
\end{equation}
here $\Delta \phi_{i,j}=[(\theta_i-\beta)^2/2-\psi(\theta_i)]-[(\theta_j-\beta)^2/2-\psi(\theta_j)]$ is the
Fermat potential difference between different images at angular positions
$\theta_i$ and $\theta_j$. The source is located at angular $\beta$.
$\psi$ is the effective two-dimensional lensing potential, which is the integral
of the Weyl potential along the line of sight.
The MG effect will enter the Fermat potential, and thus give the anomaly of the time-delay compared to the GR case.

Identifying the lensed GW+EM events is a great challenge. In principle, the crucial part is a cross-confirming procedure in both GW and EM domains. A single detection in one domain should trigger an associated search in the data from the other one. For example, if GW data analysis provides a pair of events suspected of being lensed, it should trigger a search for lensed (repeated) EM transients in the sky location strip of the GW source. Conversely, if a lensed kilonova event is observed in a large survey telescope, this should trigger confirmation searches in the GW signal database for coherent waveforms and time delay, which are synergistic with EM signal~\cite{Liao:2017ioi}.
After getting the measurements of the source and lens galaxy redshifts ($z_s$, $z_l$) as well as
the spectroscopic measurement of the stellar velocity dispersion ($\sigma_v$) of the lens galaxy,
we can calculate the time-delay of the two images for different source positions.
The lens equation here should be
\begin{equation}
\vec{\beta}=\vec{\theta}-\vec{\alpha}(\vec{\theta})\,,
\label{eq:lens}
\end{equation}
where $\alpha$ is the reduced deflection angle. From Eq.~\ref{eq:alphab}, the reduced deflection angle for a point mass lens model is
\begin{equation}
\vec{\alpha}_p(\vec{\xi})=\frac{D_{ls}}{D_s}\hat{\vec{\alpha}}_p=\frac{D_{ls}}{D_s}\frac{4GM}{c^2}\vec{\xi} \mathcal{I}(\vec{\xi})\,.
\end{equation}
Here, $\xi=D_l \theta$ is the impact factor.
However, in a real galaxy lens model, the deflection angle should be obtained by summing
the contribution of all the mass elements. Then the reduced angle is
\begin{equation}
\vec{\alpha}(\vec{\xi})=\frac{4G}{c^2}\frac{D_{ls}}{D_s}\int\Theta(\vec{\xi}')(\vec{\xi}-\vec{\xi}')\mathcal{I}(|\vec{\xi}-\vec{\xi}'|)d^2\xi'\,.
\label{eq:alphaMG}
\end{equation}
Here we denote the surface mass density of the lens galaxy as $\Theta(\vec{\xi})=\int \rho(\xi,z)dz$. For a specific lens model, given any source position, we can solve the lens equation to obtain the solution of image position from Eqs.~\ref{eq:lens} and~\ref{eq:alphaMG}. The angular diameter distance of the source and lens are restricted to the background $\Lambda$CDM model. Using the relation between the effective lensing potential and the reduced deflection angle $\nabla \psi=\alpha$, we can calculate the Fermat potential and then the final time-delay from Eq.~\ref{eq:dt}. The main goal of this work is to compare the time-delay predictions between the GR and MG theory thus in order to detect the MG effect.

The uncertainty of this lensed GW+EM system is dominated by the
measurement uncertainty of velocity dispersions.
With the current lensing image reconstruction technique, for lensed quasars, the uncertainty is of order $\mathcal{O}(3\%)$~\cite{Liao:2014cka,Suyu:2016qxx}.
The velocity dispersion measurement uncertainty is of order $\mathcal{O}(10\%)$~\cite{Jee:2014uxa,Birrer:2018vtm}.
As demonstrated by ~\cite{Liao:2017ioi} and ~\cite{Wei:2017emo}, the uncertainty of image reconstruction of EM
counterparts of GW events ($\sigma_{\theta}$), such as short Gamma-ray burst and kilonovae, is of order $\mathcal{O}(0.6\%)$.
Due to the better control of PSF in the EM counterpart of the GW event, the uncertainty of the velocity dispersion measurement ($\sigma_{\sigma_v^2}$) could be
improved to $\mathcal{O}(5\%)$~\cite{Liao:2017ioi,Wei:2017emo}.
In addition, other matter structures
along the line of sight might contribute an extra systematic uncertainty ($\sigma_{\rm LOS}$) around the $1\%$ level~\cite{Suyu:2009by,Suyu:2016qxx}.
As we have demonstrated above, the uncertainty of time-delay measurement from the lensed GW events can be neglected.
Thus the total uncertainty translated to the final time delay is straightforward, $\sigma_{\Delta t}=\sqrt{\sigma^2_{\sigma_{v}^2}+\sigma_\theta^2+\sigma_{\rm LOS}^2}$, which is
dominated by the measurement of stellar velocity dispersions. Then we can calculate how much the deviations of MG from GR can propagate to the anomalies of the time delay and compare them with the systematic errors to see whether we can distinguish them.

\begin{figure}
\includegraphics[width=0.45\textwidth]{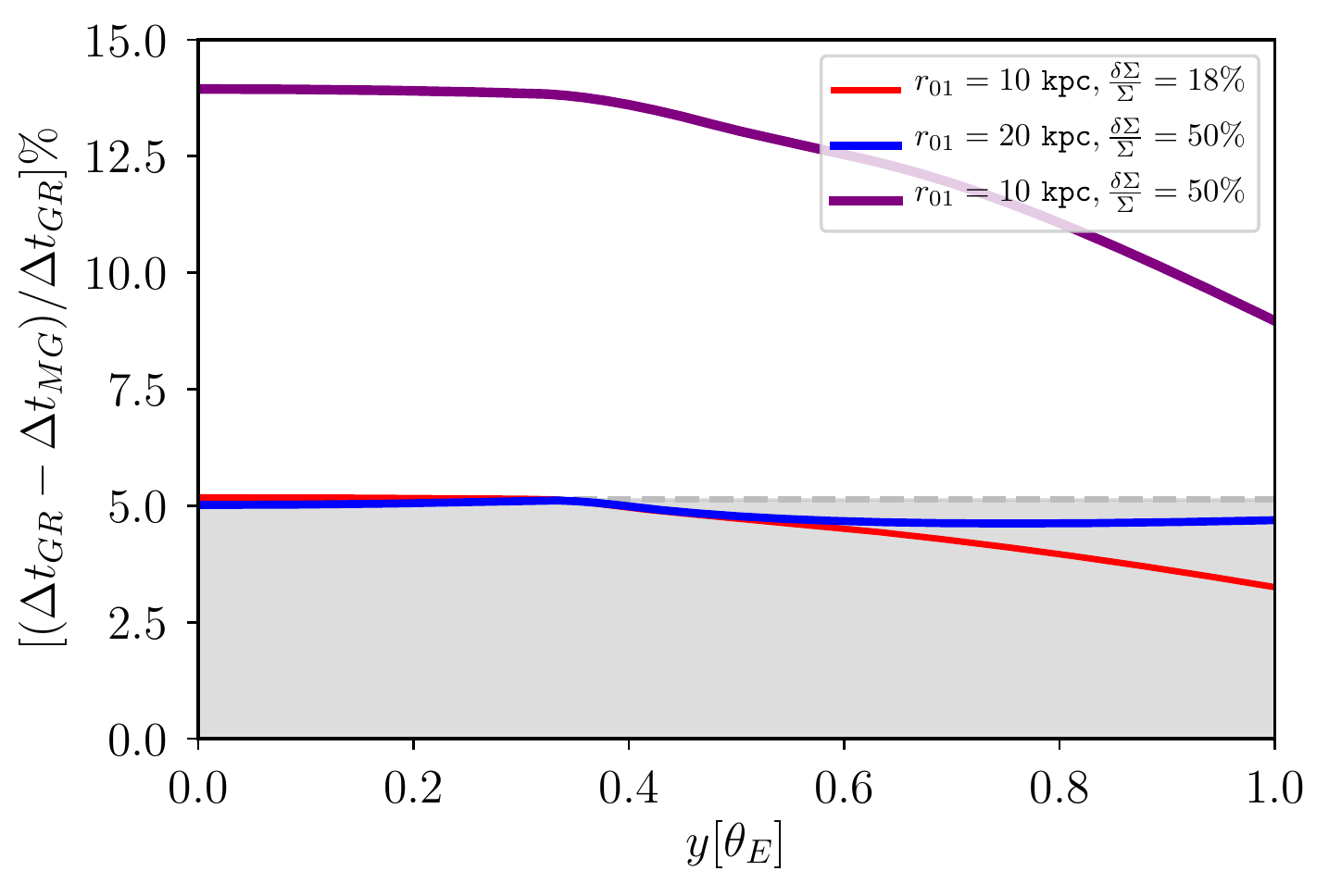}
\caption{\textbf{Time-delay difference between GR and modified gravity for source position inside the Einstein radius.} The shaded area represents the total systematic uncertainty in time delay. Only the modified gravity signals, which are upon this uncertainty can be detected. $y=\beta/\theta_E$ is the source position in the unit of Einstein radius ($\theta_E$). Here we choose a typical strong lensing system with an SIS galaxy lens and $z_l=0.3$, $z_s=1$, $\sigma_v=300$ km/s. The modified gravity effect is parameterized by the screening scale $r_{01}$, the width of the screening shell $w_{01}$ and the amplitude $\delta \Sigma/\Sigma$. Several different parameter values of $r_{01}$ and $\delta \Sigma/\Sigma$ are shown in the legend and we fix the width $w_{01}=1.2$ kpc as a thin shell model, which is not sensitive on the scale ($\sim\mathcal{O}(10)$ kpc) we are concerned with .}
\label{fig:work1}
\end{figure}

In this paper, we want to verify the viability of using lensing time-delay to distinguish the MG effect. We focus on the error propagation from the MG to the anomaly of time delay. Thus here we calculate only one typical strong lensing system and choose a simple lens galaxy model.
As an example, we specify a typical strong gravitational lensing system according to the
sample collected by Cao {\it et al.}~\cite{Cao:2015qja} with $z_l=0.3$,
$z_s=1$, and $\sigma_v=300$ km/s. The lens galaxy mass profile is set to be
the SIS model with a $10$ kpc radius. The background model is
fixed to the $\Lambda$CDM model with $H_0=67.8$ km s$^{-1}$ Mpc$^{-1}$ and $\Omega_m=0.3$. Since the SIS lens model produces double images only when the
source position is inside the Einstein radius, we plot the results with $\beta<\theta_E$.
As shown in Fig.~\ref{fig:work1}, when the MG starts to take effect above the $r_{01}=10$ ($20$) kpc scale, a
$\alpha_1/\alpha_2=18\%$ ($50\%$) deviation can be discovered by one typical strong
lens (red/blue curve). Comparing the purple curve ($r_{01}=10$ kpc, $\alpha_1/\alpha_2=50\%$) with the former two, we
can conclude that a smaller screening scale or a larger  deviation from GR can result in the easier detection of the MG signal.


\section{Consistency relationship between times delay and multiple images} \label{sec:test}
Using the time delay to test gravity is limited by large uncertainty of the measurement of stellar velocity dispersions. Remember that we have two observations, that is, the time-delay and multiple images. These two observations both reflect the information of the lensing potential. In Section~\ref{sec:timedelay} we only use the time delay from GW to infer the 
MG effect embedded in the lensing potential. This is a straightforward way. Here, we can use the image information as a supplement to help improve the constraints of MG parameters. The basic idea is that, for a scale dependent modification of the lensing potential, the response to the MG effect from GW and EM domain should be significantly different compared to the GR case. This inspires us that we can check the consistency relationship between the time delay and the multiple images under the MG framework.
Similarly, for simplicity, we also assume an SIS lens model to verify the viability of our methodology.
Hence, we can use this consistency relation, Eq.(\ref{eq:dtGR}), to test GR and also to successfully avoid the dispersion measurement
\begin{equation}
\Delta t_{i,j}=\frac{1+z_l}{2c}\frac{D_l D_s}{D_{ls}}(\theta_i^2-\theta_j^2)\,.
\label{eq:dtGR}
\end{equation}
If we assume that the ``true'' Universe is governed by the law of MG, applying a typical lensed GW+EM system as before, we can obtain the time delay and positions of multiple images.
Then we can compare the time delay between the
``true'' one with the one calculated from the $\Delta t--\theta$ relation under the GR+SIS assumption.
If the difference exceeds the systematic uncertainty, the MG signal is detectable.
Here the systematic uncertainties are only in the image position measurements and the mass along the line-of-sight contributions, $\sigma_{\Delta t}=\sqrt{\sigma_\theta^2+\sigma_{\rm LOS}^2}$.

In this paper, we are looking for some scale dependent MG effect, namely the deviation of Sigma from unity is a function of scales. For this type of MG effect, mathematically, we do not need the measurement of dynamical mass as long as we have two (or more) different tracers of the lensing mass, for example, GW time delay and associated EM lensing image. The reasons are the followings. Take the simplest case as an example, GR plus SIS lens modelling. Eq.~(\ref{eq:dtGR}) predicts a specific prediction of the EM lensing image (position angle) and the GW time delay. The benefit of this relation is that it does not involve the Weyl potential. However, any modifications to gravity or lens mass modelling will break this unique consistency relation. In this section, we are aiming to use this consistency relationship as the smoking gun of MG gravity. For simplicity, we do not consider the degeneracy between the MG effect and lens mass modelling, ie., insisting on the SIS model. Of course, the dynamical mass input can always help us and gives an additional contribution to constrain MG parameters. This is exactly what we did in the section~\ref{sec:timedelay}. Although ignoring this degeneracy is aggressive and dangerous, as in the first trial, we want to quantify the most optimistic limit of this approach. 

\begin{figure}
\includegraphics[width=0.45\textwidth]{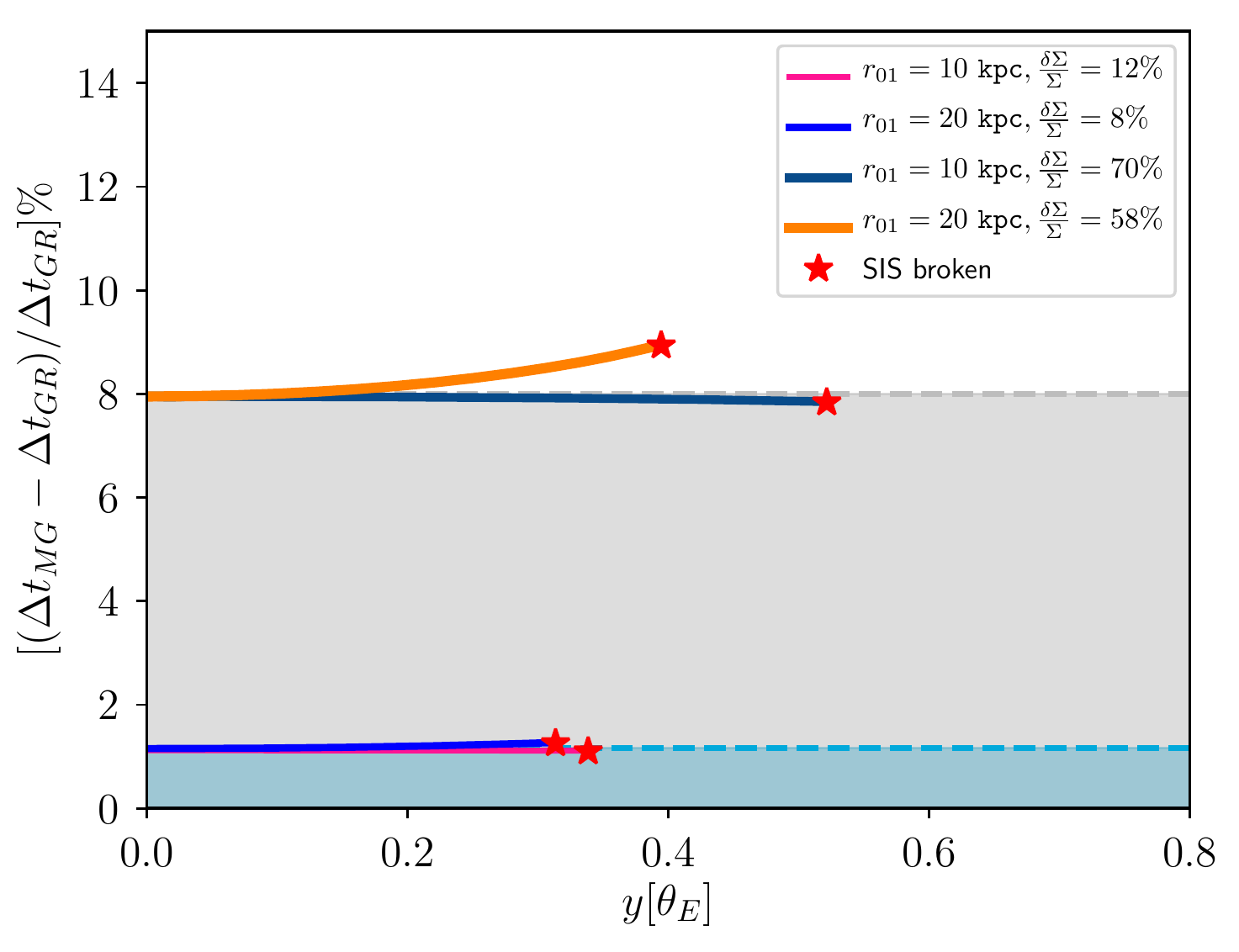}
\caption{\textbf{Consistency relationship between time-delay and multiple images.} The parameter values are similar to those in Fig.~\ref{fig:work1}. The blue shaded area represents the total
systematic uncertainty of the $\Delta t--\theta$ relationship. The gray shaded area denotes the time-delay differences under the GR+SIS assumption,
if $H_0$ changes from $67$ to $73$ km s$^{-1}$ Mpc$^{-1}$.
Here the stars at the end of each curve denote for the boundary outside which the $\Delta t--\theta$ relationship breaks, because the image lays outside
of the lens galaxy mass distribution.}
\label{fig:work2}
\end{figure}

Fig.~\ref{fig:work2} shows the result using both time-delay and image positions.
We can see that the time-delay difference is more sensitive to the magnitude than the screening scale.
For a typical strongly lensed GW+EM event, an $8\%$ deviation of MG from GR can be detected.
Furthermore, because the time delay can also arise due to the background expansion,
we need to quantify the degeneracy between MG effect and Hubble expansion, 
To do so, we choose two very different values of present Hubble constant, $H_0$, namely $\sim67$ km s$^{-1}$ Mpc$^{-1}$ from Planck~\cite{Ade:2015xua} and $\sim73$ km s$^{-1}$ Mpc$^{-1}$ from SNe Ia~\cite{Riess:2016jrr}.
If the time-delay difference is larger than those from the $H_0$ variation, we can conclude that this effect is unlikely to be due to the uncertainty of the cosmic expansion measurement.
We need to have significant modifications to gravity on the kiloparsec scale in order to fully account for the current tension in the Hubble parameter.
As seen from Fig.~\ref{fig:work2}, the latter can contribute an $8\%$ uncertainty at most. Thus, if we have a $60\%$ deviation from GR, we would have strong confidence
to claim the deviation from $\Delta t--\theta$ consistency relation is caused by MG rather than the inaccurate measurement of $H_0$.

\section{Discussion} \label{sec:discussion}
To measure the GW time-delay, we need to record the time sequence when the maximum peak arrives.
Compared with the standard sirens method~\cite{Schutz:1986gp}, our approach evades the complicated wave form calculation.
Furthermore, due to the much smaller uncertainties of time-delay and image position measurements,  the strongly lensed GW+EM system can give
a better estimation of lensing mass compared with the lensed quasar. 

In section~\ref{sec:timedelay} we propagate all the errors (including velocity dispersion) to the time-delay and calculate the difference of time-delay between GR and MG and derive an error estimation (18\%) of the MG detection. This method does not rely on the specific lens model. Both the time delay and lensing image response to the same Weyl potential, but the relation between them varies among different models of gravity. In section~\ref{sec:test}, we propose to test the simplest relation, Eq.~(\ref{eq:dtGR}), which is valid in the GR+SIS assumption. The goodness of this relation is that it avoids the measurement of velocity dispersion of the lens galaxy, which is one of the major uncertainties in this approach. However, this assumption is aggressive because it breaks when either the GR or SIS assumption is invalid. The reason we still test this relationship is that we want to quantify the upper limit of this approach in the optimistic lens modelling assumption.

The PPN parameter is constrained up to the order of $10\%$ by Collett {\it et al.} using the ESO 325-G004.
We should note that the uncertainty of the dynamical mass they derived is about $4\%$ which is smaller than the one we adopted, conservatively ($5\%$).
Instead of assuming a constant PPN parameter, we use a scale dependent $\Sigma$ function to parameterize the MG effect.
Since the gravitational lensing, in principle, is the redistribution of the relativistic radiation, the spatial derivative is essential.
We argue that the constant PPN parameter is oversimplified in the lensing data analysis.
Furthermore, the traditional method of using the optical strong lensing system is highly limited by the spectroscopic measurement. Take the ESO 325-G004 lens as an example, its image data were obtained by HST at 2006, but the IFU spectroscopic data from Multi Unit Spectroscopic Explorer were obtained nearly 10 yr later. More importantly, to obtain a high quality spectroscopic data, we are limited to the the low redshift lens, around $z_l\sim 0.03$. Even with such high quality data, we can only test gravity up to the $3$ kpc scale with the $10\%$ accuracy~\cite{Collett:2018gpf}. By using the method proposed in section~\ref{sec:test}, although its assumption is aggressive, it is completely independent from the spectroscopic measurement. Hence, it can be used at higher redshift. The sample numbers can be greatly increased. If some significant modifications to gravity, say a $>10\%$ effect on Weyl potential, are detected, they can trigger further studies on the gravity test.  

In our calculation, we set the galaxy radius to be $10$ kpc. The SIS model $\rho(r)=\frac{\sigma_v^2}{2\pi Gr^2}$ with $\sigma_v=300$ km/s, gives the total mass $6.57\times 10^{11} M_{\odot}$.
Most of the galaxy radii are about $0.5-50$ kpc, which correspond to the total mass $10^{10}-10^{12}M_{\odot}$.
Thus the galaxy we set here is a typical one.

Though some MG theories would be excluded on some specific scales. For example the generalized Brans--Dicke (GBD) models predict a $\Sigma$ which is strongly constrained to be close to unity in models with universal coupling to different matter species.  However, the deviation from GR on any scale 
should attract our attention, especially for the galactic scale on which the GR are poorly constrained. A measurement of $\Sigma\neq 1$ would rule out all universally coupled GBD theories, such as the f(R),  chameleon, symmetron, and dilaton models. In this paper, we propose the idea of using the strongly lensed multimessenger to distinguish any deviation of the phenomenological parameter from the GR prediction.

We would emphasize here that our approach is more general and robust compared to the traditional method. The upcoming strongly lensed multimessenger system, such as the GW+EM events, may not be well aligned with the observer. Thus the traditional way to utilize the Einstein radius is impractical. In this paper, we take the SIS lens model as an example to demonstrate how much the MG effect can be detected. For a more general lens model, such as singular isothermal ellipsoid (SIE), the calculation is straightforward. Our approach is not limited by the specific source type or the lens model. 

Although GW lensing is an exciting and relatively novel topic, it faces many challenges. Up to now, we do not observe any lensed GW event. There are many challenges facing the realistic GW lensing data analysis. One of the major issue is finding GW lensed sets in the time stream from a GW detector. Considering the confusion of background (unresolved) sources and variable noise/duty cycles of a detector, this identification procedure is far from straightforward. If a set can be done, there is another challenge of finding a unique galactic host in the large localization area. If a GW signal has been strongly magnified, the `effective redshift' inferred (assuming no magnification) will be very different from the host redshift, making identification more sophisticated. And depending on source-lens-observer setup, it is possible for signals to sometimes get demagnified, lowering the GW amplitude and potentially rendering them undetectable. We cannot tell more about this issue because the corresponding pipelines have not been built. However, in this paper, we proposed a new methodology that can be adopted in the future test of gravity.

\acknowledgments
We thank Ran Li for helpful discussions and comments. 
The authors also thank the anonymous referee for his/her helpful comments and suggestions. 
TY and BH are supported by the Beijing Normal University Grant under the reference No. 312232102 and by the National Natural Science Foundation of China Grants No. 11690023 and No. 11653003.  
TY is also supported by China Postdoctoral Science Foundation under Grants No. 2017M620662.
BH is also partially supported by the Chinese National Youth Thousand Talents Program under the reference No. 110532102 and the Fundamental Research Funds for the Central Universities under the reference No.310421107. BW acknowledges the support by NNSFC No.11835009. RGC was supported  by the National Natural Science Foundation of China  Grants No.11690022, No.11375247, No.11435006, and No.11647601, and by the Strategic Priority Research Program of CAS Grant No.XDB23030100 and by the Key Research Program of Frontier Sciences of CAS.


\bibliography{ref}

\end{document}